\documentclass[aps,a4paper,twocolumn,floatfix,showkeys,
groupaddress]{revtex4}

\usepackage{bm,amsmath,amssymb,amsfonts}

\usepackage[dvips]{graphicx}
\usepackage[colorlinks=true,citecolor=green,urlcolor=red,filecolor=blue]{hyperref}
\usepackage{color}
\definecolor{mycolor}{rgb}{0.80,0.10,0.10}
\begin{document}

\title{\textcolor{mycolor}
{Tunable caging of excitation in decorated Lieb-ladder geometry with long range connectivity}}  

\author{Atanu Nandy}
\email{atanunandy1989@gmail.com}
\affiliation{Department of Physics, Acharya Prafulla Chandra College, New Barrackpore, Kolkata
West Bengal-700 131, India}

\begin{abstract}
Controlled Aharonov-Bohm caging of wave train is reported in a quasi-one dimensional version of Lieb geometry with next nearest neighbor hopping integral within the tight-binding framework. This longer wavelength fluctuation is considered by incorporating periodic, quasi-periodic or fractal kind of geometry inside the skeleton of the original network. This invites exotic eigenspectrum displaying a 
distribution of \textit{flat band states}. Also a subtle modulation of external magnetic flux leads to a comprehensive control over those non-resonant modes. Real space renormalization group method provides
us an exact analytical prescription for the study of such tunable imprisonment of excitation. The non-trivial tunability of external agent is important as well as challenging in the context of experimental perspective.
\end{abstract}
\maketitle
\section{Introduction}
\label{intro}
Recent exciting headway in experimental condensed matter physics helps us to emulate several quantum mechanical phenomena in a quite tunable environment. 
This unprecedented advancement in fabrication technique provides a scope for direct visualization of different
 theoretically proposed phenomena like localization of excitation in low dimensional networks~\cite{seba1,seba2}. That is why creation of so called artificial systems for the simulation of complex many-body systems containing additional degree of freedom has grabbed considerable scientific impact~\cite{rodrigo}.
Moreover, scientific communities have already addressed the celebration of sixty
 years of the pioneering work of Anderson~\cite{anderson}. The absence of diffusion
 of wave packet in the random disorder environment is well known. In fact this now becomes a general prescription in diverse topics of condensed matter physics starting from optical lattice 
 of ultra cold atoms~\cite{bloch} to the acoustics, 
 wave guide arrays~\cite{leder} or in micro cavities having 
 exciton-polaritons~\cite{kim}. Unlike the case of Anderson localization (AL), the concept of compact localized states (CLS)~\cite{dias}-\cite{flach5} in several one or two dimensional periodic or non-periodic structures has attracted the spot light of fundamental research. The journey started nearly thirty years ago approximately from Sutherland~\cite{suther}.

This unconventional non-diffusive progress of wave has generated significant attention because of its contribution to various novel physical phenomena in
strongly correlated system, such as unconventional Anderson localization~\cite{goda,shukla}, 
Hall ferromagnetism~\cite{tasaki,derzhko}, high-temperature superconductivity~\cite{kau},
and superfluidity~\cite{torma}, to name a few.
Moreover, this study has kept scientists intrigued since it offers a suitable platform to investigate
several phenomena that are linked with the information of quantum physics together with the topological effect including fractional quantum hall effect~\cite{gong} and flat band ferromagnetism~\cite{aoki}. For these CLS, the diminishing envelope of the wave train beyond finite size \textit{characteristics trapping cell} implies extremely low group velocity due to the divergent effective mass tensor. This means that the particle behaves like a super heavy such that it cannot move. The vanishing curvature of the $E-k$ plot corresponding to such momentum independent self-localized states are generally caused by the destructive nature of the quantum interference occurred by multiple quantum dots and the local spatial symmetries involved with the underlying structure. Hence these are also called as flat band states.

In general, occurrence of dispersionless flat band can be classified into two categories depending on their stability with respect to the application of magnetic perturbation. 
In particular, the type of geometries discussed by Mielke~\cite{mie} 
and Tasaki~\cite{tasaki} cannot contain flat bands for finite magnetic flux. Whereas, the other type of lattices e.g., Lieb lattice~\cite{lieb1}, there exists macroscopically degenerate flat band even in the presence of flux. In fact, the non dispersive band is completely insensitive to the applied external perturbation. As it is well known that 
the inherent topology of the line-centered square lattice (also known as the Lieb lattice) induces interesting spectral properties such as the macroscopically degenerated zero-energy flat band, the Dirac cone in the low-
energy spectrum, and the typical Hofstadter-type spectrum in a magnetic field. Moreover, Lieb geometry is one of the most prominent candidate useful for  magnetism. The spectral divergence of the zero-energy flat band provides that platform.

In this manuscript, inspired by all the experimental realizations of Aharaonov-Bohm caging, we study a quasi-one dimensional Lieb-ladder network within the tight-binding formalism. The phenomenon of imprisonment of wave train is studied when the next nearest neighbor (NNN) connection term is added to the Hamiltonian. Interesting modulation of self-trapping of excitation is also studied in details when the NNN connectivity is `decorated' by either magnetic flux or some quasi-periodic, fractal kind of objects.

As a second motivation we have analyzed an Aharonov-Bohm interferometer model made in the form of a quasi-one dimensional Lieb geometry to study the flux controlled localization aspects. 
It is needless to mention that this flux controlled caging is a subset of widely used phenomena Aharonov-Bohm caging~\cite{vidal} and this has
been experimentally verified in recent times~\cite{seba1,seba2}.
However, when an electron traverses a closed loop that traps a finite magnetic flux $\Phi$, its wave function picks up a phase factor. This simple sentence is at the the core of the pioneering Aharonov-Bohm (AB)
 effect~\cite{aharonov}-\cite{markus3} which has led to a substantial research in the standard
 AB interferometry that dominated the fundamental physics, both theoretical and experimental perspective, in the mesoscopic scale over the past few decades~\cite{yacoby1}-\cite{kubo2}.
It is to be noted that the current experiments by Yamamoto \textit{et al.}~\cite{yamamoto} has stimulated more experiments on quantum transmission in AB interferometers~\cite{aharony6}. Also the previously mentioned theoretical model studies have also played 
 an important part in studying the elementary characteristics of the electronic states and coherent conductance in quantum networks in the mesoscopic dimensions~\cite{kubo2}. The recent advancement in the fabrication and lithography processes have opened up the possibility to make a tailor-made geometry with the aid of quantum dots (QD) or Bose–Einstein condensates (BEC). It is needless to mention that this has provoked a substantial content of theoretical research even in model quantum networks
 with a complex topological character~\cite{silva,souza}.
 
In this article, highly motivated by the ongoing scenario of theory and experiments in AB interferometry, we investigate the spectral and the transmission properties of a model quantum network in which
diamond shaped Aharonov-Bohm interferometers are arranged in the form of a quasi-one dimensional Lieb ladder geometry. 
Such diamond-based interferometer
models have previously been analyzed as the minimal 
prototypes of bipartite networks having nodes with different 
coordination numbers, and representing a family of itinerant geometrically
 frustrated electronic systems~\cite{lope}. There are other studies which include the
 problem of imprisonment of excitation under the influence of spin-orbit interaction~\cite{ber}, 
 a flux-induced semiconducting behavior~\cite{ac}, quantum level engineering for AB cages~\cite{movi} or, 
 as models of spin filters~\cite{cohen}.
 
In what follows we demonstrate our findings. Sec.~\ref{modellieb} discusses the basic quasi-one dimensional Lieb ladder network in respect of energy band and transmittivity. In Sec.~\ref{diamondlieb} we have incorporated a next nearest neighbor connectivity by inserting a rhombic loop inside the unit cell and discussed the flux sensitive localization. After that in Sec.~\ref{fibo} the NNN hopping is decorated by a quasiperiodic Fibonacci geometry and the distribution of \textit{self-localized} states has been studied. Sec.~\ref{frac} demonstrates the self-similar pattern of compact localized states as a function of magnetic flux. In Sec.~\ref{ifm} we have studied the Lieb Aharonov-Bohm interferometer model in respect of its electronic eigenspectrum. Finally in Sec.~\ref{close} we draw our conclusions.

\section{Model system and Hamiltonian}
\label{modellieb}
\begin{figure}[ht]
\centering
(a) \includegraphics[clip,width=7.5 cm,angle=0]{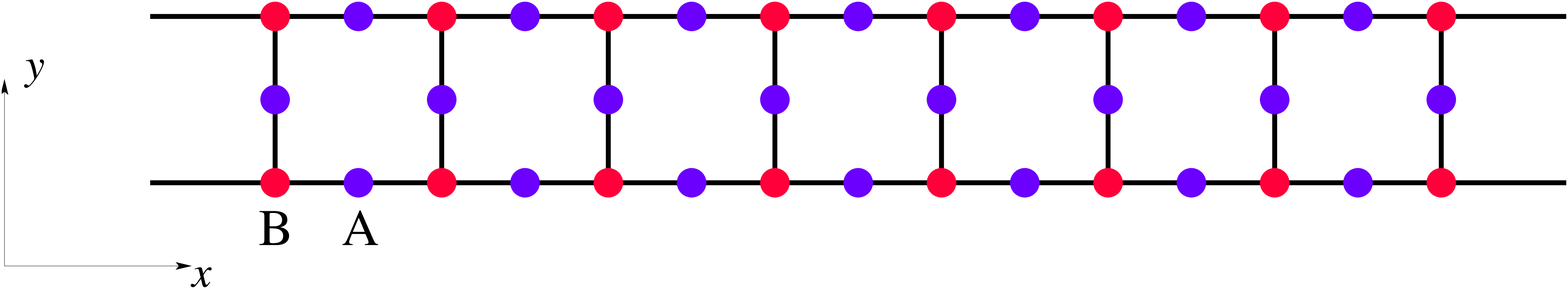}\\
(b) \includegraphics[clip,width=7.5 cm,angle=0]{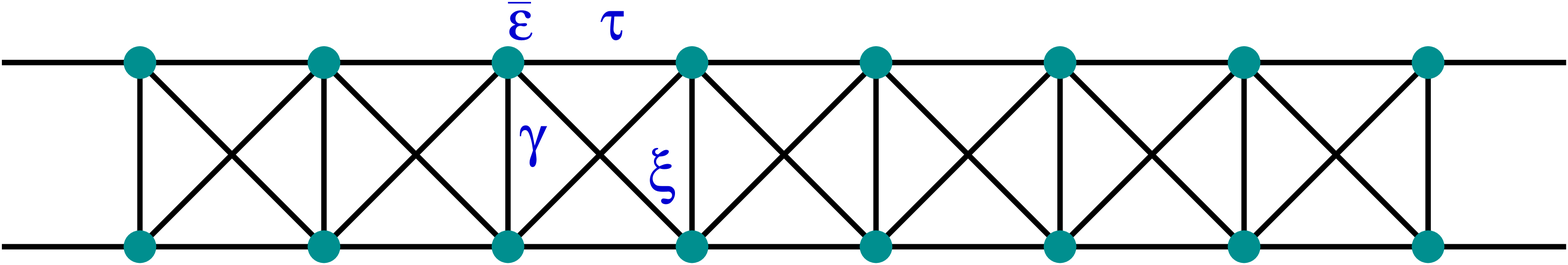}
\caption{(Color online) (a) A quasi-one dimensional Lieb ladder network with endless axial span and (b) 
the \textit{effective} two-arm ladder with \textit{renormalized} parameters.}  
\label{lattice}
\end{figure}

 We start our demonstration from the Fig.~\ref{lattice}(a) where a quasi-one dimensional version of the Lieb geometry is shown. We make a distinction between the sites (blue colored dots marked as $A$ site and red colored dots marked as $B$ sites) based on their coordination numbers.
 The
 array is modeled by the standard tight-binding Hamiltonian written in the Wannier basis, viz., 
 \begin{equation}
H=\sum_{j} \epsilon_{j} c_{j}^{\dagger} c_{j} + \sum_{\langle jk \rangle} [t_{jk} c_{j}^{\dagger} c_{k} + h.c.]
\label{hamilton}
\end{equation}
where the first term bears the potential information of the respective quantum dot location and the 
second one indicates the kinetic signature between two neighboring lattice sites.
The on-site potential of the respective sites are marked as $\epsilon_A$ and $\epsilon_B$ and the nearest neighbor overlap parameter can be assigned as $t$. Without any loss of generality, numerically the site potentials are taken as uniform (equal to zero) and the nearest neighbor hopping is also same (equal to unity) everywhere. By virtue of real space renormalization group (RSRG) technique one can easily \textit{eliminate}
the amplitude of an appropriate subset of nodes to caste the original system into an \textit{effective} two-strand ladder system with renormalized parameters as cited in the Fig.~\ref{lattice}(b). 
The decimation method can be easily implemented with the help of difference equation,
the discretized form of the Schr\"{o}dinger's equation, viz.,
\begin{equation}
(E-\epsilon_{j}) \psi_{j} = \sum_{k} t_{jk} \psi_{k}
\label{diff}
\end{equation}
This decimation provides the renormalized uniform two-leg ladder network with different parameters.
After this renormalization procedure, all the atomic sites carry identical on-site energy $\bar{\epsilon}$
and the intra-arm hopping $\tau$. The inter-arm vertical connectivity is marked as $\gamma$ as cited in the 
Fig.~\ref{lattice}(b). This decimation produces a next nearest neighbor hopping, denoted by $\xi$, which 
generates overlap between the wave functions of the two diagonally opposite atomic sites.
 The detailed forms of those parameters are given by,
\begin{eqnarray}
\bar{\epsilon} &=& \epsilon+ \frac{2 t^{2} (E-\epsilon_{1})}{\delta} \nonumber\\
\tau &=& \frac{t^{2} (E-\epsilon_{1})}{\delta} \nonumber\\
\gamma &=& \frac{2 t^{2} t_{1}}{\delta} \nonumber\\
\xi &=& \frac{t^{2} t_{1}}{\delta}
\label{decimation1}
\end{eqnarray}
where $\epsilon_{1} = \epsilon+t^2/(E-\epsilon)$, $t_{1} = t^2/(E-\epsilon)$ and $\delta = [(E-\epsilon_{1})^{2} - t_{1}^{2}]$.
With the above \textit{renormalized} parameters and by virtue of RSRG approach, one can trivially compute the electronic density of states (DOS) $\rho(E)$ for this quasi-one dimensional Lieb strip as a function of the energy of the incoming projectile by using the standard expression, viz.,
\begin{equation}
\rho (E) = - \left( \frac{1}{N \pi} \right) Im [Tr G(E)]
\label{dens}
\end{equation}
Here $G(E) = [E-H+i \Delta]^{-1}$ is the usual green's function and $\Delta$ is
 the imaginary part of the energy, reasonably small enough, added for the numerical 
 evaluation of DOS. $N$ denotes the total number of atomic sites present in the 
 system and `Tr' is the trace of the green's function. 
\begin{figure}[ht]
\centering
(a) \includegraphics[clip,width=6 cm,angle=0]{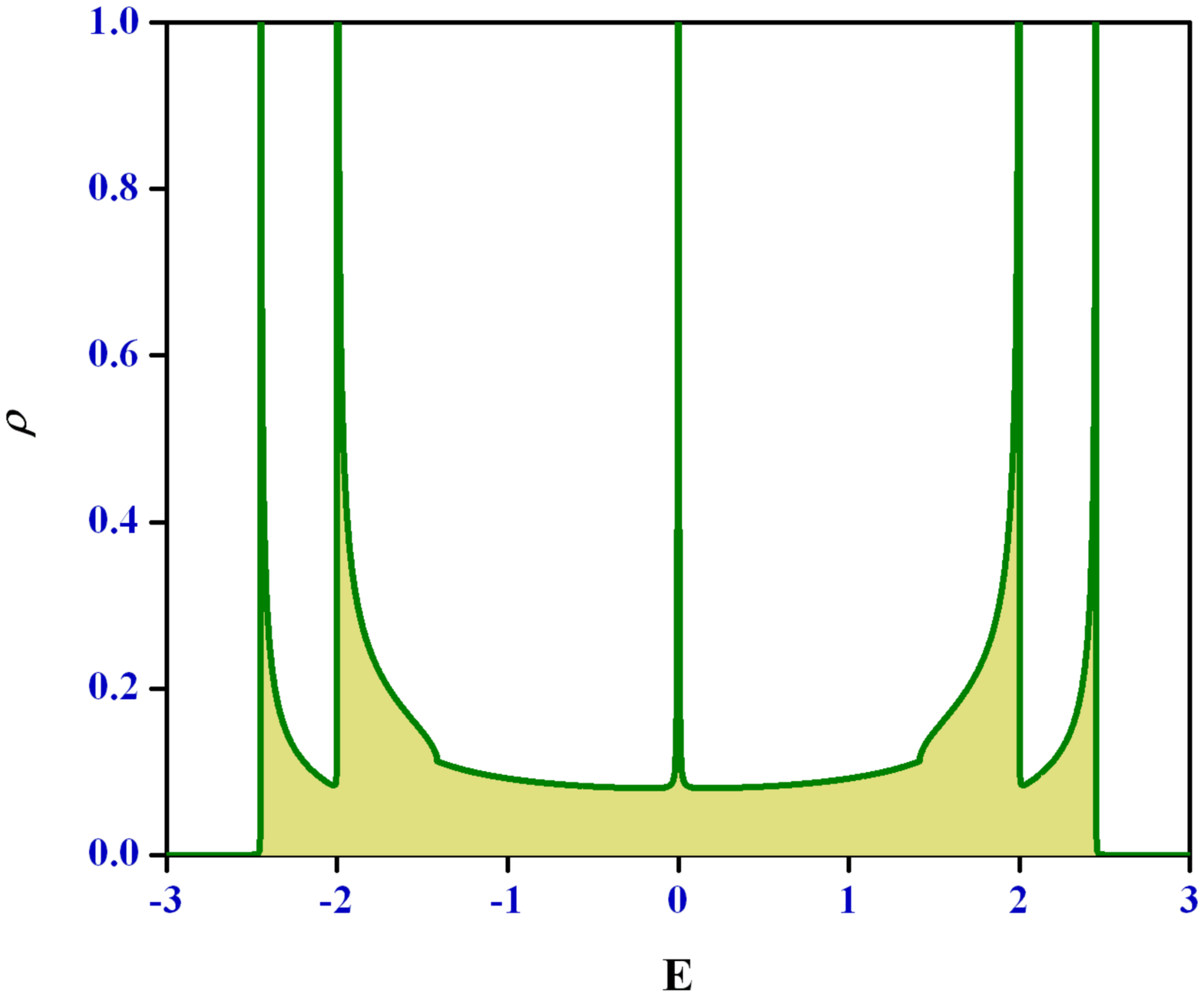}\\
(b) \includegraphics[clip,width=6 cm,angle=0]{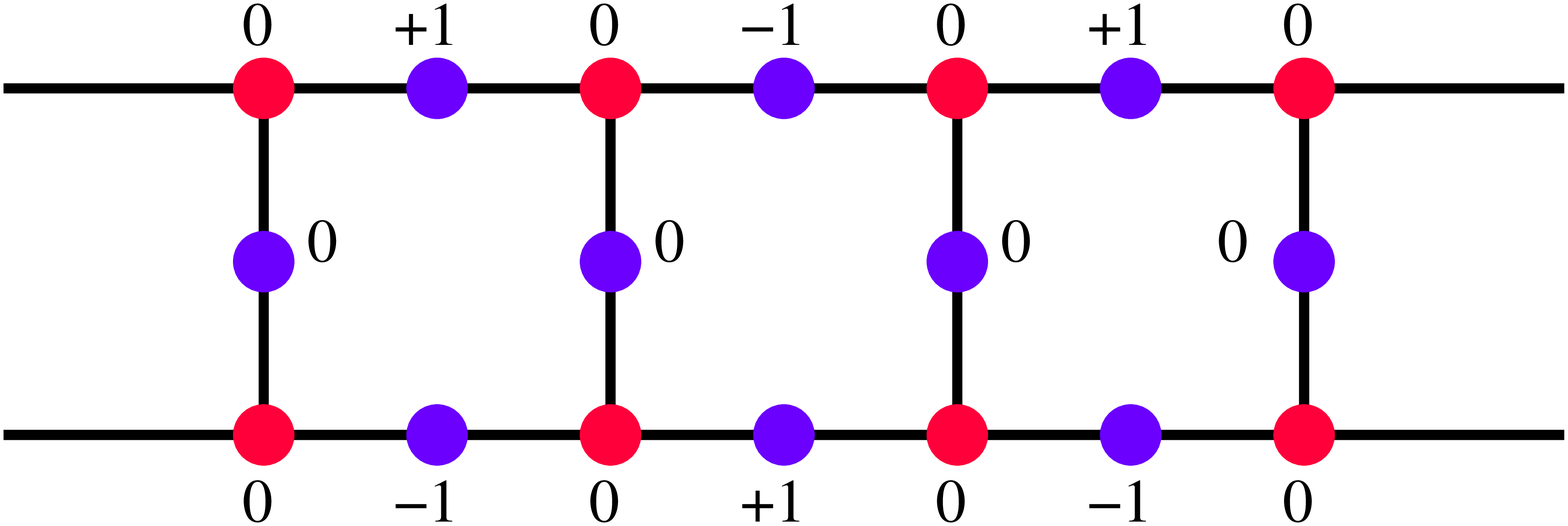}\\
(c) \includegraphics[clip,width=6 cm,angle=0]{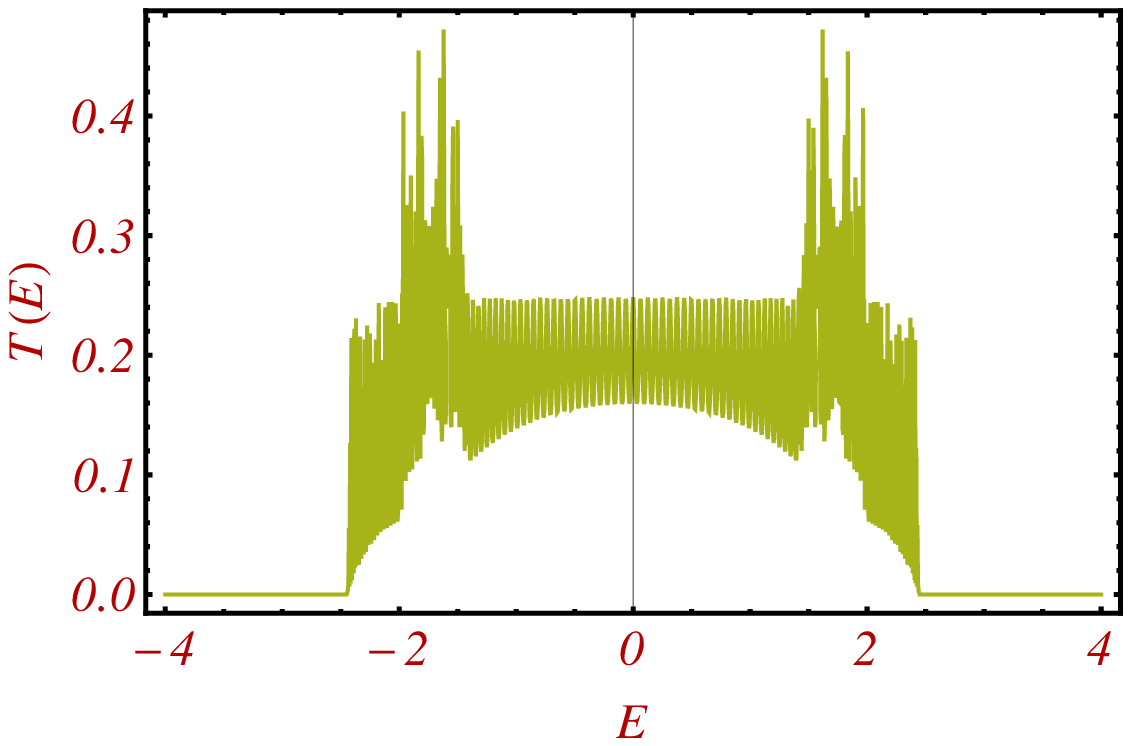}
\caption{(Color online) (a) Plot of density of eigenstates as a function of energy $E$ for quasi-one
dimensional Lieb-ladder geometry, (b) denotes the amplitude distribution profile for $E=0$ and 
(c)variation of transmittance with energy.}  
\label{lieb1d}
\end{figure}
\subsection{Density of eigenstates and transport}
In Fig.~\ref{lieb1d}(a) the variation of DOS is presented as a function of energy where we see the presence of the absolutely continuous Bloch bands populated by extended eigenfunctions. We have checked that for any energy belonging to the resonant band, the overlap parameter keeps on non-decaying behavior and that is a signature of the state being delocalized. At the band center $(E = 0)$, the central spike confirms the existence of momentum independent flat band state which is an inherent signature of the Lieb geometry. 
The spectral divergence corresponding to the zero energy mode comes from the
vanishing group velocity of the wave packet as $\rho \propto \int v_{g}^{-1} d\mathbf{k}$. With the aid of 
difference equation one can obtain the distribution of amplitude for such self-localized eigenstate.
The non-vanishing amplitudes are pinned at the intermediate sites as shown in Fig.~\ref{lieb1d}(b)
and one such \textit{characteristic trapping island} is isolated from the other by a 
distinct physical boundary formed by the sites with zero amplitude as a result of destructive quantum interference.
The dispersionless nature of the central band 
is responsible for anomalous behavior in the transport and optical properties.
The construction of this state definitely resembles the essence of a \textit{molecular state} which is
spatially quenched within a finite size cluster of atomic sites.
 The analogous wave function does not present any evolution dynamics beyond the 
\textit{trapping cell}. Extremely low mobility of the wave train is the key factor for the dispersionless signature of the state.
But here we should point out that since the compact localized state, thus formed, lies inside the continuum zone
of extended states, here the hopping integral never dies out for $E=0$. Hence, one should observe non-zero
transport for that particular mode. The localization character can be 
prominently viewed in presence of any perturbation when the spectrum shows central gap around $E=0$, if any.

To corroborate the above findings related to the spectral landscape we now present a precise discussion
 to elucidate the electronic transmission characteristics for this quasi-one dimensional system. For this analysis we have considered a finite-sized underlying network. Now the ladder-like system needs to be clamped in between two pairs of semi-infinite periodic leads with the corresponding parameters. One can then adopt the standard green's function approach~\cite{low1,low2} and compute the same for the composite system (lead-system-lead). The transmission probability~\cite{supriyo1}-\cite{paro3}
 can be written in terms of this green's function including the self-energy term as,
\begin{equation}
\tau_{ij}=Tr[\Gamma_{i}G_{i}^{r}\Gamma_{j}G_{i}^{a}]
\label{transmit}
\end{equation}
Here the terms $\Gamma_{i}$ and $\Gamma_{j}$ respectively 
denote the connection of the network with the $i$-th and $j$-th leads and $G$'s are the retarded and
advanced Green's functions of the system.
The result is demonstrated in the Fig.~\ref{lieb1d}(c). It describes a wide resonant window for which
 we have obtained ballistic transport. The existence of Bloch-like eigenfunctions for this wide range of Fermi energy is solely responsible for this high transmission behavior. The conducting nature of the spectral density is basically reflected in this transmission plot.
\subsection{Band dispersion}
To study the energy-momentum relation of this periodic system
 we will cast the original Hamiltonian in terms of wave vector $k$ by virtue of the following expression,
\begin{equation}
H = \sum_{k} \psi^{\dagger}_{k} \mathcal{H} (\bm{k}) \psi_{k}
\label{ktransform}
\end{equation}
Using this relation, the Hamiltonian matrix in $k$-space reads as,
\begin{equation}
\mathcal{H}(\bm{k})=
\left[ \begin{array}{ccccc}
\epsilon & t & 0 & t(1+e^{-ika}) & 0\\
t & \epsilon & t & 0 & 0\\
0 & t & \epsilon & 0 & t(1+e^{-ika})\\
t(1+e^{ika}) & 0 & 0 & \epsilon & 0\\
0 & 0 & t(1+e^{ika}) & 0 & \epsilon
\end{array}
\right]
\end{equation}
\begin{figure}[ht]
\centering
\includegraphics[clip,width=6 cm,angle=0]{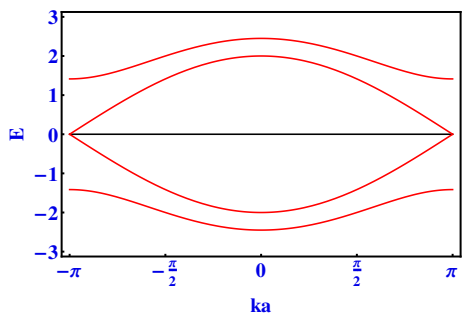}
\caption{(Color online) Band dispersion diagram of a quasi-one dimensional Lieb-ladder network showing the 
central \textit{flat band} and other two pairs of \textit{dispersive} bands.}  
\label{liebdisp}
\end{figure}
The straightforward diagonalization of the above matrix reveals the entire band picture of the Lieb-ladder network as presented in Fig.~\ref{liebdisp}. It clearly shows one momentum insensitive 
non-dispersive band at $E=0$ with absolutely zero curvature and two pairs of Bloch bands carrying dispersive signature at $E= \pm \sqrt{2(1+ \cos ka)}$
and $E= \pm \sqrt{2(2+ \cos ka)}$.
The central flat band state confirms the existence of robust type of \textit{molecular state}. 

\section{Diamond-Lieb network}
\label{diamondlieb}
\begin{figure}[ht]
\centering
\includegraphics[clip,width=7.5 cm,angle=0]{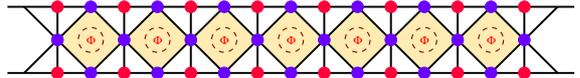}
\caption{(Color online) A quasi-one dimensional array of Lieb-ladder geometry with 
next nearest neighbor (NNN) hopping term incorporated by a diamond loop threaded by uniform magnetic 
flux $\Phi$.}  
\label{geo2}
\end{figure}
In the previous description presented so far, the off-diagonal element, i.e., the hopping parameter is taken to be restricted within the nearest neighboring atomic sites only within the tight-binding formulation. 
We now consider the same quasi-one dimensional Lieb-ladder geometry with next nearest 
neighbor (NNN) hopping integral taken into consideration between the $A$ types of sites as cited in the Fig.~\ref{geo2}. With the inclusion of longer range connectivity the entire periodic geometry turns out to be quasi-one dimensional Lieb ladder
 with a rhombic geometry embedded inside the skeleton. This additional overlap parameter introduces another closed loop within each unit cell where the impact of application of magnetic perturbation may be examined in details.

\begin{figure}[ht]
\centering
(a) \includegraphics[clip,width=6 cm,angle=0]{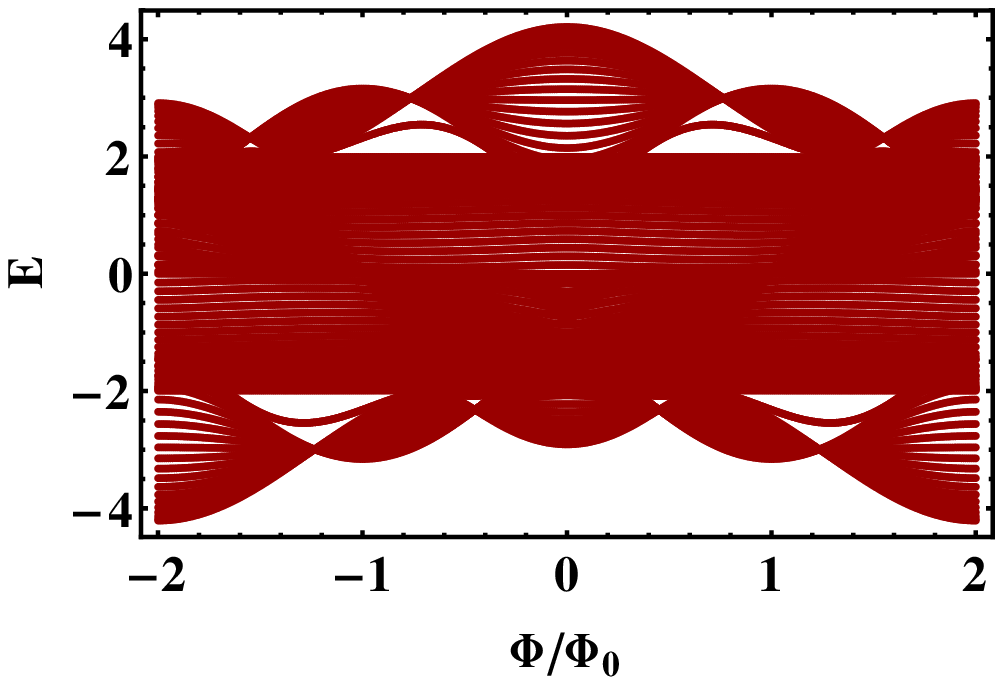}\\
(b) \includegraphics[clip,width=6 cm,angle=0]{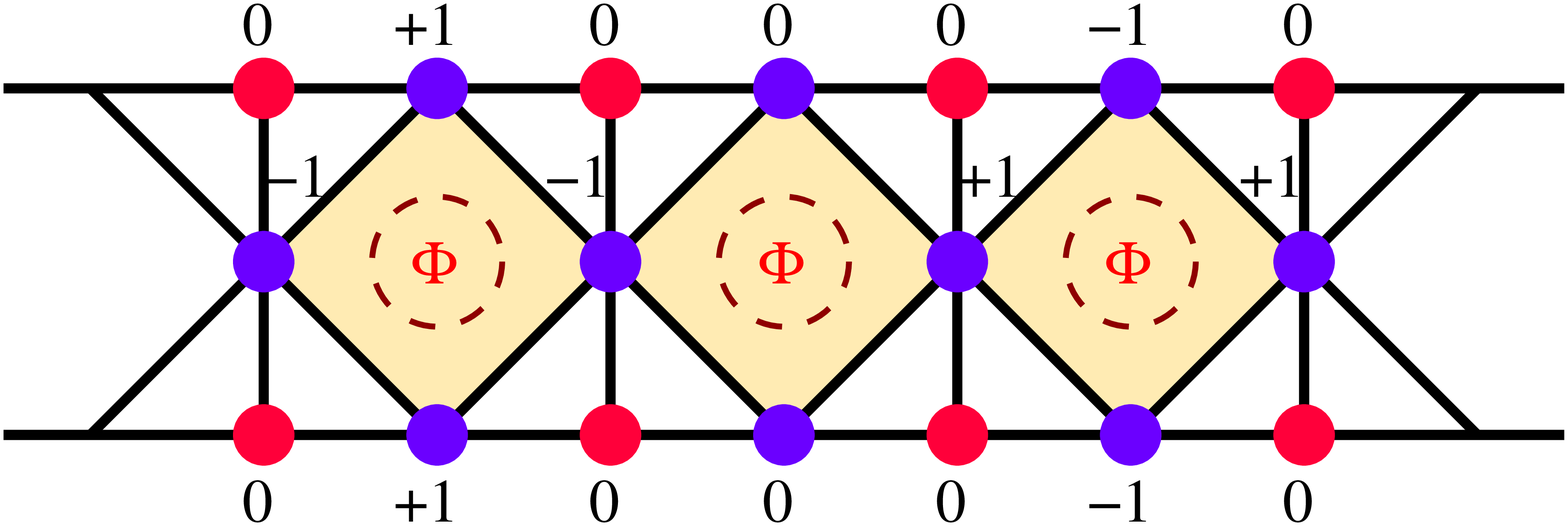}
\caption{(Color online) (a) Presentation of allowed eigenspectrum as a function of magnetic flux
for diamond-Lieb network and (b) amplitude profile corresponding to the energy $E = \epsilon - 2 t \cos \Theta$.}  
\label{spec}
\end{figure}
Before presenting the numerical results and discussion it is necessary to mention that uniform magnetic perturbation may also be applied within each rhombic plaquette. This can be feasible by an appropriate choice of the gauge. This can introduce additional externally tunable parameter which may lead to interesting 
band engineering. 
This flux tunable localization of excitation will be discussed in the subsequent subsection.

\subsection{Allowed eigenspectrum as a function of flux}
Now we analyze the impact of uniform magnetic perturbation on the sustainability of the 
self-localized states. The magnetic flux is applied inside each embedded rhombic plaquette.
\begin{figure*}[ht]
\centering
(a) \includegraphics[clip,width=5 cm,angle=0]{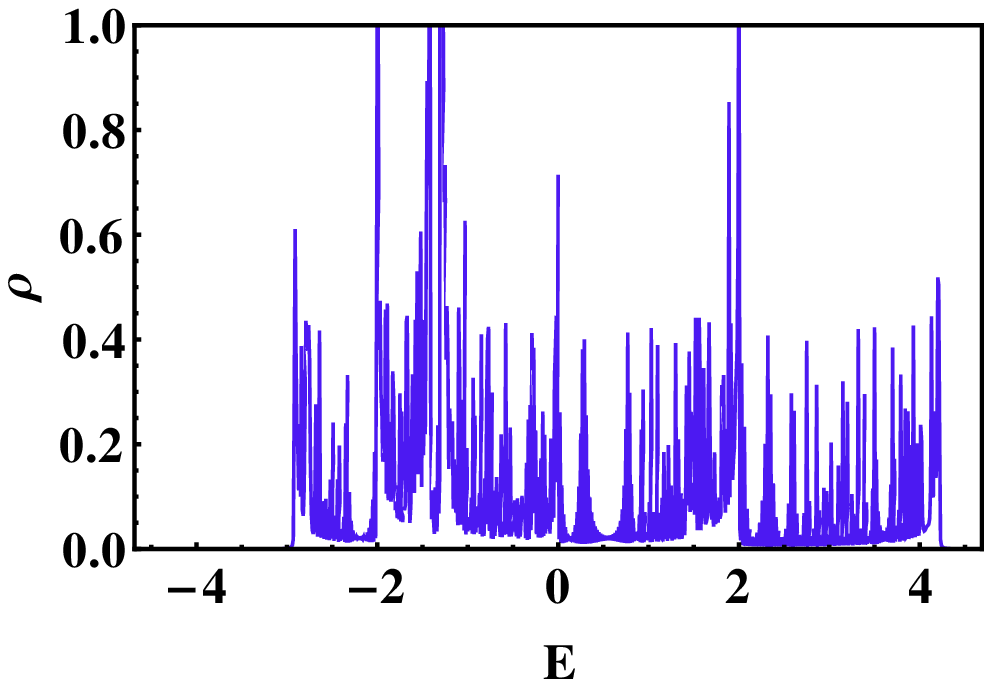}
(b) \includegraphics[clip,width=5 cm,angle=0]{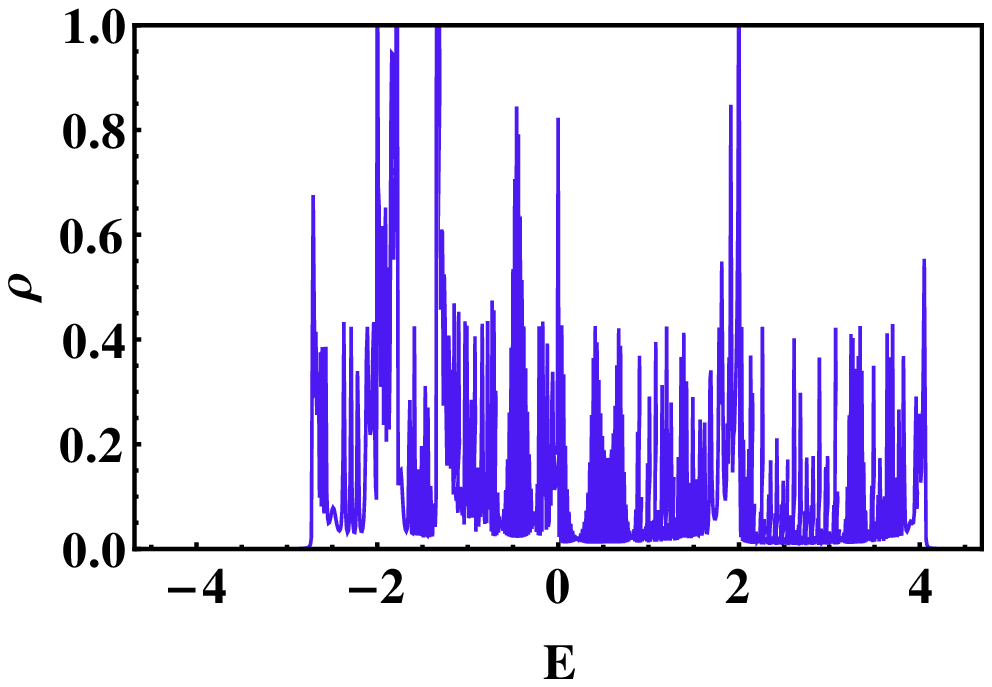}
(c) \includegraphics[clip,width=5 cm,angle=0]{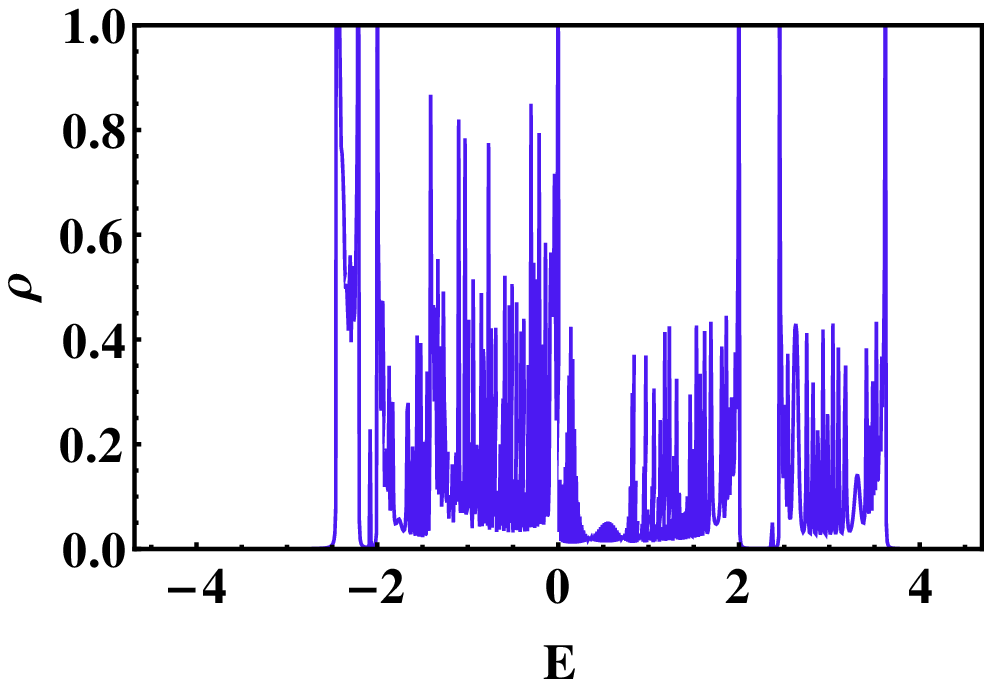}
\caption{(Color online) Variation of density of states $\rho(E)$ as a function of energy $E$ of the excitation.
The external magnetic flux values are respectively (a) $\Phi = 0$, (b) $\Phi = \Phi_0 /4$ and (c) $\Phi = \Phi_0 /2$.}  
\label{diadosphi}
\end{figure*}
As a result of this application of magnetic flux, the time reversal symmetry is broken (at least locally) along the arm of the rhombic plaquette.
This is considered by introducing a Peierls' phase factor associated with the hopping integral, viz., $t \rightarrow t e^{i \Theta}$, where, $\Theta = 2 \pi \Phi/4 \Phi_{0}$ and $\Phi_0 = hc/e$ is termed as fundamental flux quantum.
The resultant nature of quantum interference
happened due to multiple quantum dots is the ultimate determining factor
for the sustainability of the \textit{self-localized} modes after applying the
 perturbation.
 Here we have evaluated the allowed eigenspectrum (Fig.~\ref{spec}(a)) with respect to the 
 applied flux for this flux included quasi-one dimensional diamond-Lieb geometry. The spectrum is
 inevitably flux periodic. Multiple band crossings, formation of several minibands and thus merging of each other
 are seen in this quasi-continuous pattern.

Here we should give emphasis on a pertinent issue. Fig.~\ref{spec}(b) shows a consistent demonstration of amplitude profile (satisfying the difference equation) for energy $E = \epsilon - 2 t \cos \Theta$, 
$\epsilon$ being the uniform potential energy everywhere. One non-vanishing cluster is again isolated from the other by a physical barrier formed by the sites with zero amplitude as a direct consequence of phase cancellation at those nodes. This immediately tells us that the incoming electron coming with this particular value of energy will be localized inside the trapping island. But now the energy eigenvalue is sensible to the applied flux which is an external agency.
The central
motivation behind the
  application of this external parameter is that if possible, we may invite a comprehensive tunability of such bound states solely by manipulating the applied flux. We do not need to disturb any internal parameter of the system, instead one can, in principle, control the band engineering externally by a suitable choice of flux.
  The external perturbation can be tuned \textit{continuously} satisfying the eigenvalue equation to control the position of the caged state.
  
\subsection{Density of states profile}
For the completeness of the analysis, we have 
\begin{figure}[ht]
\centering
\includegraphics[clip,width=6 cm,angle=0]{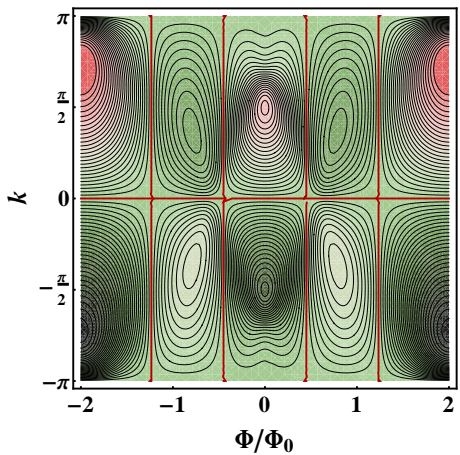}
\caption{(Color online) $k-\Phi$ diagram showing different group velocity contours for electron moving 
in diamond embedded Lieb geometry. The red lines mark the zero group velocity of the wave packet. 
These red contours act as border lines showing a continuous change of $v_g$ with respect to flux.}  
\label{vel}
\end{figure}
computed the variation of density of states profile as a function of energy of the incoming projectile for this quasi-one dimensional lattice with longer wavelength fluctuation using the standard green's function technique both in the absence and presence of external perturbation.
The variation with respect to the energy 
of the incoming projectile for different values of magnetic flux is shown in the Fig.~\ref{diadosphi}. The applied flux values are respectively $\Phi = 0$, $\Phi = \Phi_0 /4$ and $\Phi = \Phi_0 /2$. All the variations are plotted for system size $N = 753$. As it is evident from the plots that there are different absolutely continuous subbands populated by extended kind of eigenfunctions. The existence of such dispersive modes is expected because of the inherent translational periodicity of the geometry. 
We have examined that for any mode belonging to the continuum zones the hopping integral shows oscillatory behavior which confirms the signature of the resonant modes. It is needless to say that the intricate nature of the DOS is highly sensitive on the external perturbation.
Also the density of states plots as well as the allowed eigenspectrum support the existence of flux dependent 
caged state as discussed in the previous section.

\subsection{Band engineering}
In presence of uniform magnetic flux one can easily express the Hamiltonian in the k-space language. The 
 diagonalization of this matrix will give the band dispersion as a function of flux. 
 In this quasi-one dimensional diamond Lieb geometry we have got that, there are two flux independent dispersive bands $E = \pm \sqrt{2(1+\cos ka)}$ and three other flux sensible resonant bands. 
Therefore we should highlight a very pertinent issue here. For the last three flux dependent bands, one can easily control the group velocity of the wave train as well as the effective mass (equivalently the mobility) of the particle by tuning the external source of perturbation. This non-trivial manipulation of the internal parameters of the system with the aid of flux makes this aspect of band engineering more challenging as well as interesting indeed.

Before going to detailed discussion, it is important to be noted that, when an electron moves around a closed loop that traps a magnetic flux, the wave function picks up a phase related to the magnetic vector potential, viz., $\psi = \psi_0 e^{i \oint \mathbf{A}.d\mathbf{r}}$. 
The magnetic flux here plays an equivalent role as the wave vector~\cite{gefen}. One can thus think of a $k-\Phi/\Phi_0$ diagram which is equivalent to a typical $k_{x}-k_{y}$ diagram for electrons traveling in a two-dimensional periodic lattice. The ``Brillouin zone'' equivalents are expected to show up, across which variations 
of the group velocity will take place. This is precisely shown in the Fig.~\ref{vel}. 
In this plot, every contour presented corresponds to a definite value (positive or negative) of the group velocity of the wave packet. The red lines are the contours with zero mobility. Hence they are the equivalents of the boundaries of the Brillouin zone across which the group
 velocity reverts its sign if one moves parallel to the $\Phi$-axis at any fixed
 value of the wave vector $k$, or vice versa.
This essentially signifies that, we can, in principle,
make an electron accelerate (or retard) without manipulating its energy by changing the
 applied  magnetic flux only. 
The vanishing group velocity contours (marked by red) indicate that the associated wavefunctions are \textit{self-localized} around finite size \textit{islands} of atomic sites, making the eigenmode a non-dispersive one. As the curvature of the band is related to the mobility of the wave packet one can conclude from the Fig.~\ref{vel} that tuning of the curvature of the dispersive band is also possible with the help of external perturbation.

\section{Lieb ladder with quasiperiodic next nearest neighbor interaction}
\label{fibo}

In the previous case the amplitude for $E=0$ will be pinned at the top and down vertices of the diamond embedded. From this standpoint we now decorate each arm of the rhombic plaquette by a finite generation quasiperioidic fibonacci kind of geometry with two different hoppings $t_x$ and $t_y$ respectively. The generation sequence for this quasiperiodic structure follows the standard inflation rule
 $X \rightarrow X Y$ and $Y \rightarrow X$. Based on this prescription regarding the anisotropy in off-diagonal term, there exists three different types of atomic sites $\alpha$ (flanked by two $X$-bonds), $\beta$ (in between $X-Y$ pair) and $\gamma$ (in between $Y-X$ pair). Here we should mention that we consider the generations with $X$ type of bond at their extremities, i.e., $G_{2n+1}$, ($n$ being integer). This is only for convenience and does not alter the result Physics-wise as we go for thermodynamic limit.
\begin{figure}[ht]
\centering
\includegraphics[clip,width=5 cm,angle=0]{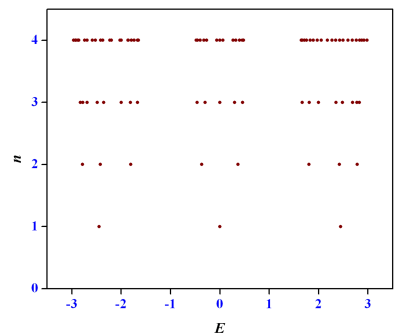}
\caption{(Color online) Distribution of self-localized modes showing a typical three-subband pattern for large
enough generation.}  
\label{root}
\end{figure}

Hence if we start with a odd generation Fibonacci segment that decorates each arm of the diamond, then one can decimate the chain $n$-times by employing the RSRG method to get back the original diamond structure with renormalized parameters. The recursive flows of the parameters are governed by the following equations, viz.,
\begin{eqnarray}
\epsilon_{\alpha}(n+1) &=& \epsilon_{\alpha}(n)+ \frac{t_{x}^2(n)}{\Delta(n)} [2E - (\epsilon_{\beta}(n)+\epsilon_{\gamma}(n))] \nonumber\\
\epsilon_{\beta}(n+1) &=& \epsilon_{\alpha}(n)+ \frac{(E-\epsilon_{\beta}(n)) t_{x}^2(n)}{\Delta(n)} + \frac{t_{x}^2(n)}{(E-\epsilon_{\beta}(n))} \nonumber\\
\epsilon_{\gamma}(n+1) &=& \epsilon_{\gamma}(n)+ \frac{(E-\epsilon_{\gamma}(n)) t_{x}^2(n)}{\Delta(n)} + \frac{t_{y}^2(n)}{(E-\epsilon_{\beta}(n))} \nonumber\\
\epsilon_{C}(n+1) &=& \epsilon_{\alpha}(n)+ \frac{2 t_{x}^2(n)}{\Delta(n)} [2E - (\epsilon_{\beta}(n)+\epsilon_{\gamma}(n))] \nonumber\\
t_{x}(n+1) &=& \frac{t_{x}^2(n) t_{y}(n)}{\Delta(n)} \nonumber\\
t_y (n+1) &=& \frac{t_{x}(n) t_{y}(n)}{(E-\epsilon_{\beta}(n))}
\label{rg-fibo}
\end{eqnarray}
where $\Delta(n) = [(E-\epsilon_{\beta}(n)) (E-\epsilon_{\gamma}(n))] - t_{y}^{2}(n)$

Obviously after decimation if we want to explore the same compact localized state (at $E = \epsilon$) in this renormalized lattice, then due to the iterative procedure, on-site potential is now a complicated function of energy. And if we now extract roots from the eigenvalue equation $(E-\epsilon_{\alpha})=0$, all the roots will produce a multifractal distribution of the set of compact localized states. Obviously as we increase the generation of the fibonacci structure, in the thermodynamic limit, all the \textit{self-localized} modes exhibit a global three subband structure. The pattern is already prominent in Fig.~\ref{root}.
Each subband can be
 fine scanned in the energy scale to bring out 
the inherent self-similarity and multifractality, the hallmark of the Fibonacci
quasicrystals~\cite{tang}. The self-similarity of the spectrum
 have been checked
by going over to higher enough generations, though we refrain from showing these data to save space here. 

\section{Lieb ladder with fractal type of long range connection}
\label{frac}
\begin{figure}[ht]
\centering
\includegraphics[clip,width=7.5 cm,angle=0]{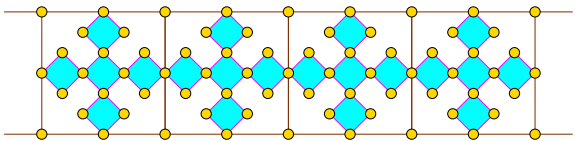}
\caption{(Color online) An infinite array of Lieb strip with long range connectivity decorated by
fractal object.}  
\label{vicmodel}
\end{figure}

We start this demonstration from the Fig.~\ref{vicmodel} where a finite generation 
of \textit{self-similar} Vicsek geometry~\cite{bp1,bp2} is grafted inside the basic Lieb motif. 
The longer range connection is here established through the aperiodic object. Also a uniform magnetic flux $\Phi$ may be applied in each small plaquette of the fractal structure. It should be appreciated that while a Lieb geometry in its basic skeleton is known to support a 
robust type of central self-localized state, 
the inclusion of fractal structure of a finite \textit{generation} in each
unit cell disturbs the translational ordering locally (though it
is maintained on a global scale in the horizontal direction) in
the transverse direction. This non-trivial competitive scenario makes the conventional methods
of obtaining the self-localized states impossible to be implemented,
especially in the thermodynamic limit. We
take the help of RSRG technique to bypass this issue and present an analytical 
formalism from which one can exactly determine the localized modes as a function of external flux.
\begin{figure}[ht]
\centering
\includegraphics[clip,width=6 cm,angle=0]{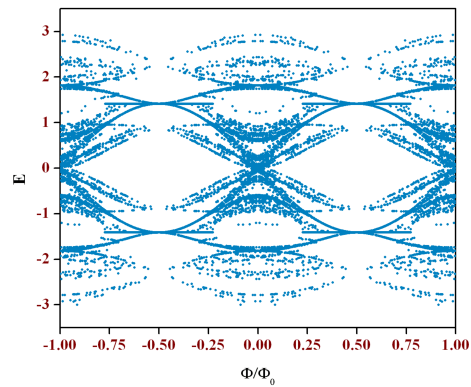}
\caption{(Color online) Distribution of self-localized states with applied flux.}  
\label{vicspec}
\end{figure}
Starting from a finite generation of scale invariant fractal network, after suitable steps of 
decimation~\cite{bp1,bp2}
 one can produce a Lieb ladder geometry with a diamond plaquette embedded into it (as discussed in the previous discussion). The renormalized potential of the top vertex of the diamond is now a complicated function of energy and flux. Therefore straightforward solving of the equation $[E- \epsilon_{A} (E, \Phi)]=0$ gives us a interesting distribution of compact localized states in the $E-\Phi$ space.

This non-trivial distribution of eigenvalues as a function of flux may be considered an equivalent dispersion relation since for an electron moving round a closed path, the magnetic flux behaves the similar physical role as that of the wave vector~\cite{gefen}. The distribution of eigenmodes
 compose an interesting miniband-like structure as a function of external perturbation. The competition between the global periodicity and the local fractal entity has a crucial impact on this spectrum.
We can continuously engineer the magnetic
 flux to engineer the imprisonment of wave train with high selectivity. Moreover, there are a number of inter-twined
 band overlap, and a quite densely packed distribution of allowed modes, forming quasi-continuous $E-\Phi$ band structure. Close observation of this eigenspectrum reveals the formation of interesting
 variants of the Hofstadter butterflies~\cite{hof}. The spectral landscape is a quantum fractal, and encoding the gaps
 with appropriate topological quantum numbers remains an open problem for such deterministic fractals.
 
Before ending this section we should put emphasis on a very pertinent point. An aperiodic fractal object is inserted in the unit cell of the periodic geometry. The \textit{self-similar} pattern of the fractal entity will have the influence on the spectrum. All such \textit{self-localized} modes are the consequences of destructive quantum interference and the geometrical configuration of the underlying system. For this class of energy eigenvalue, the spatial span of the cluster of atomic sites containing non-vanishing amplitudes increases with the generation of the fractal geometry incorporated. Hence with an appropriate choice of the RSRG index $n$, 
the onset of localization and hence the
spread of trapping island can be staggered, in space. This tunable delay of the extent of 
localization has already been studied for a wide varieties of fractal geometries~\cite{bp1,bp2,an1,an2}.
This comprehensive discussion regarding the manipulation of the geometry-induced localization makes the phenomenon of Aharonov-Bohm caging more interesting as well as challenging from the experimental point of view.

\section{Diamond-Lieb interferometer}
\label{ifm}
\begin{figure}[ht]
\centering
(a) \includegraphics[clip,width=7.5 cm,angle=0]{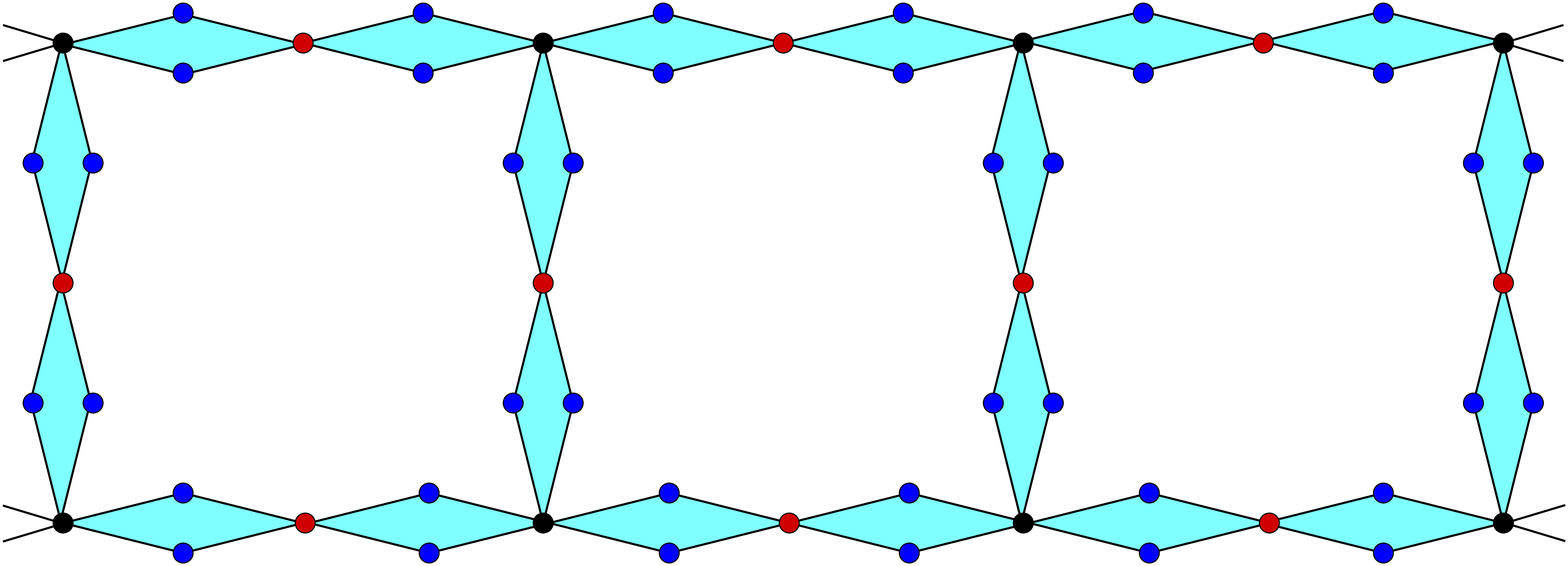}\\
(b) \includegraphics[clip,width=7.5 cm,angle=0]{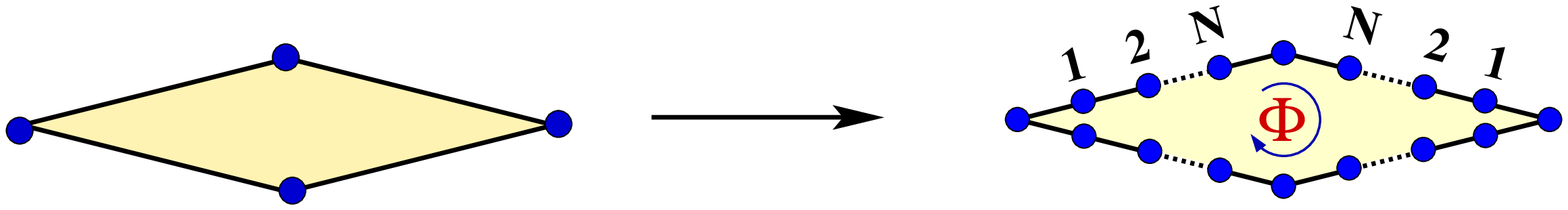}
\caption{(Color online) (a) Schematic diagram of elementary diamond-Lieb interferometer and (b) demonstrates 
the decoration of basic unit.}  
\label{ifm-model}
\end{figure}
\begin{figure*}[ht]
\centering
(a) \includegraphics[clip,width=5 cm,angle=0]{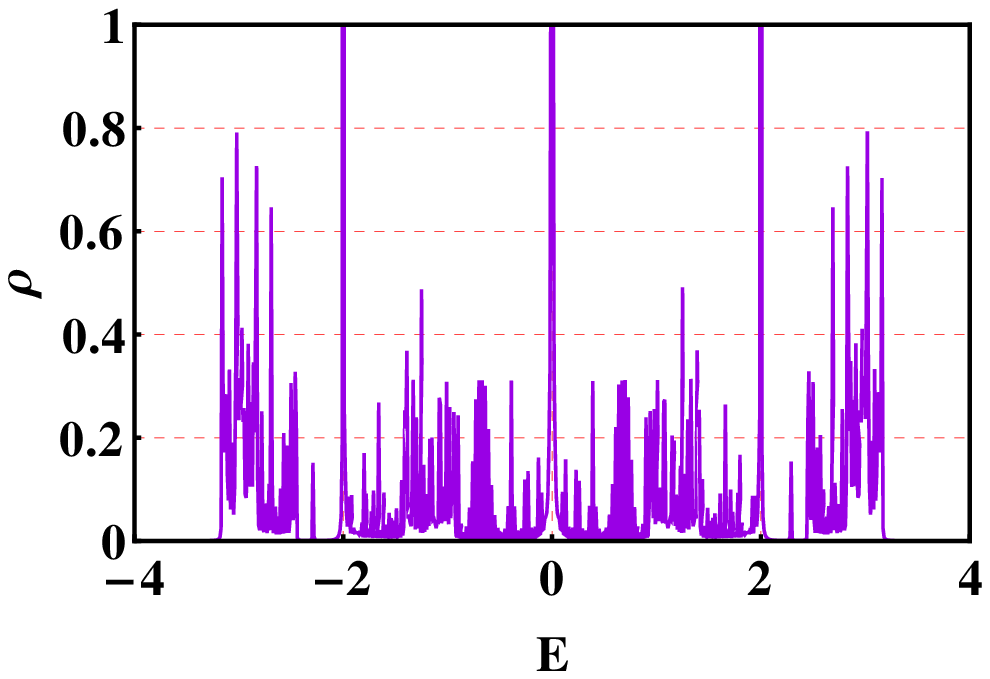}
(b) \includegraphics[clip,width=5 cm,angle=0]{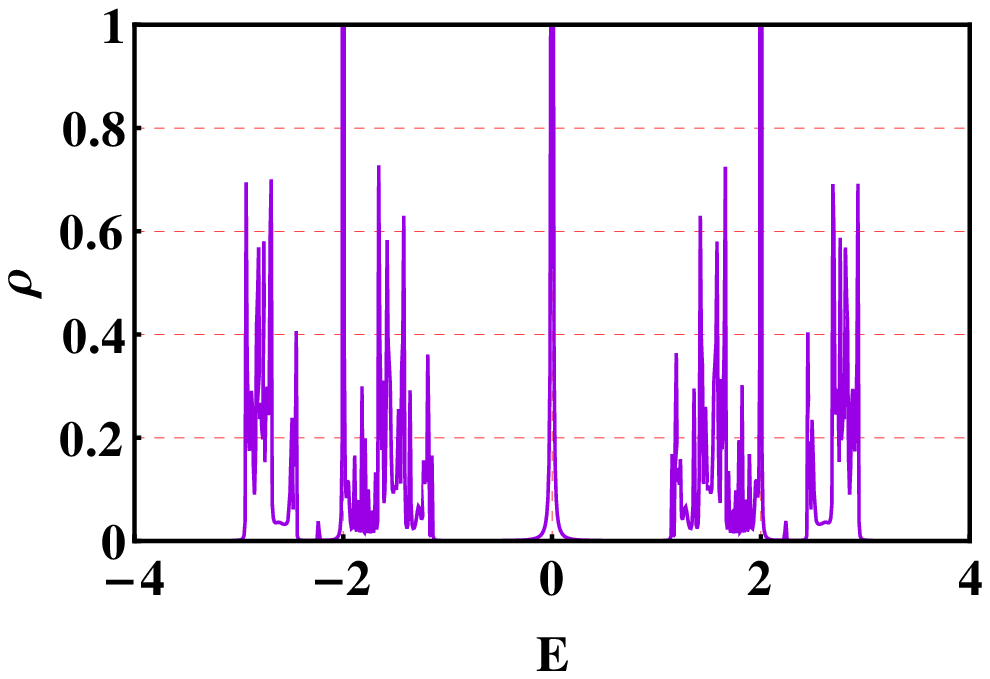}
(c) \includegraphics[clip,width=5 cm,angle=0]{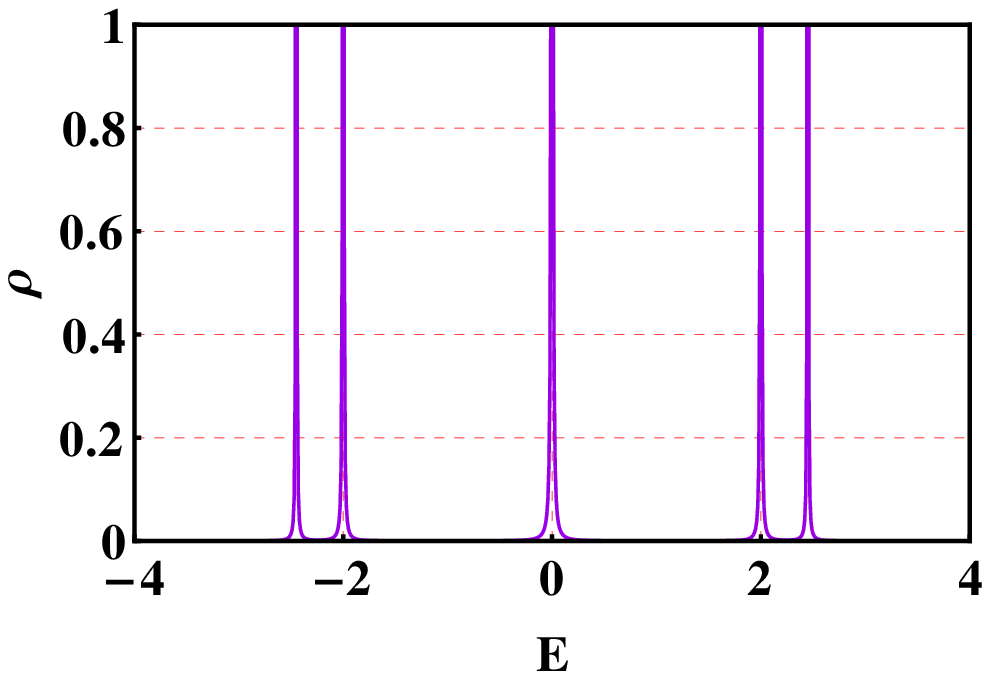}\\
(d) \includegraphics[clip,width=5 cm,angle=0]{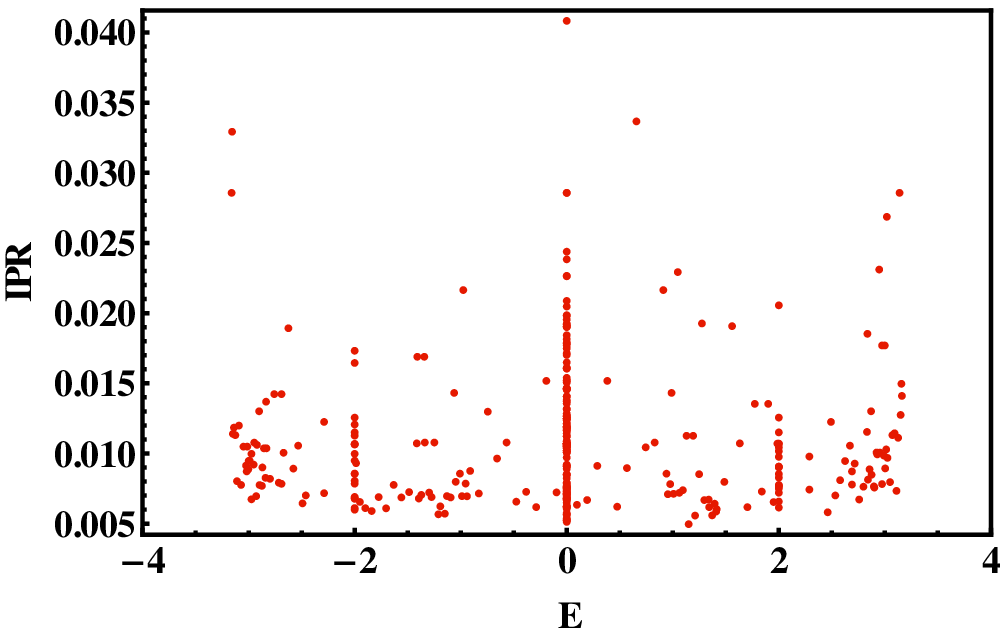}
(e) \includegraphics[clip,width=5 cm,angle=0]{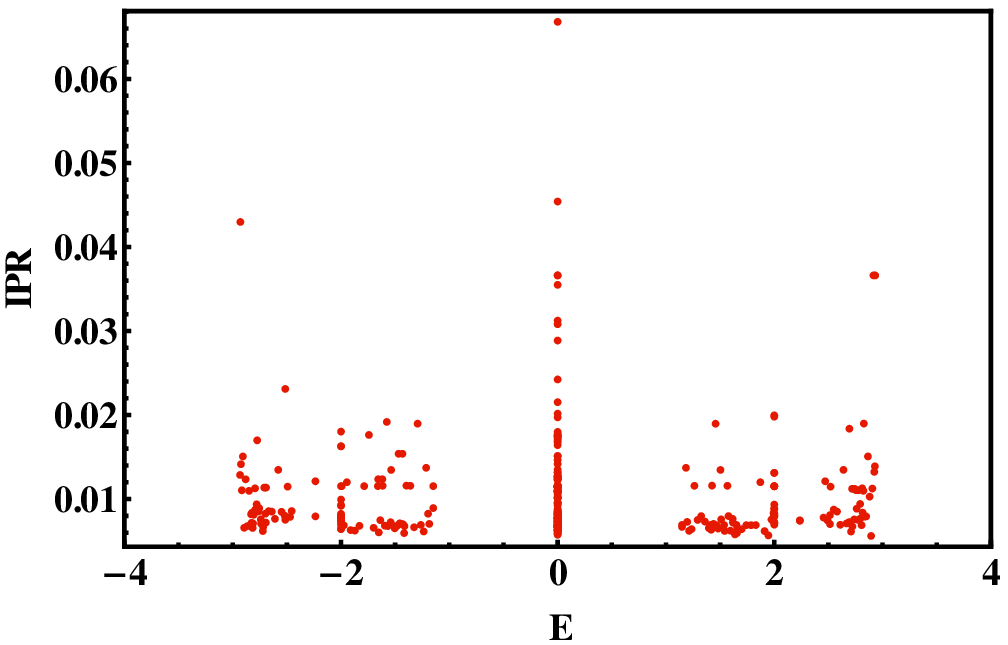}
(f) \includegraphics[clip,width=5 cm,angle=0]{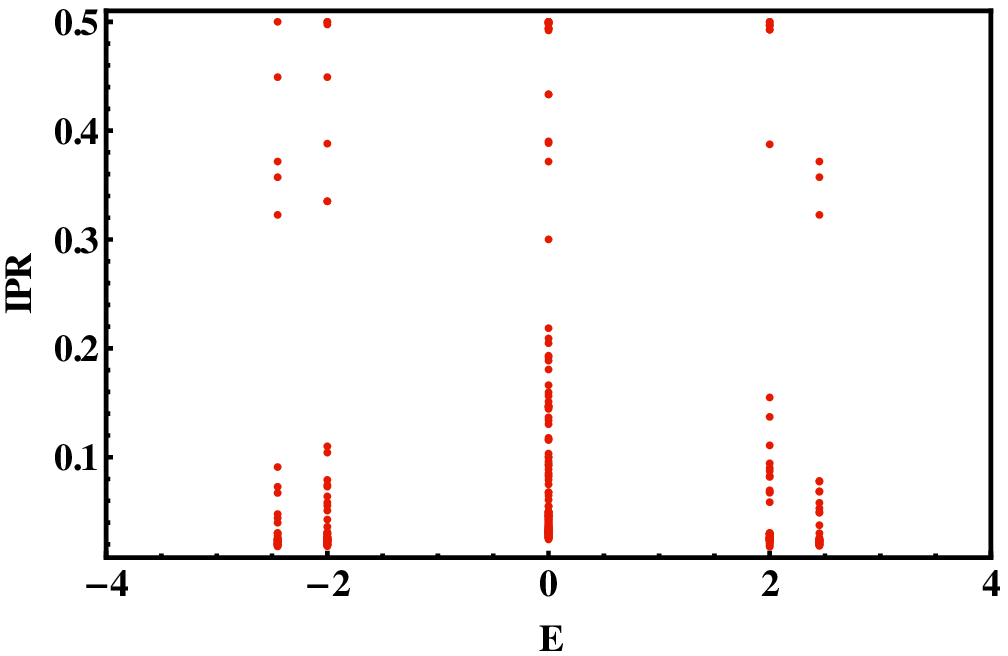}
\caption{(Color online) (Upper panel) 
Variation of density of states $\rho(E)$ as a function of energy $E$ of the excitation and
(lower panel) indicates the variation of inversion participation ratio (IPR) wth energy.
The external magnetic flux values are respectively (a) $\Phi = 0$, (b) $\Phi = \Phi_0 /4$
and (c) $\Phi = \Phi_0 /2$.}  
\label{ifm-dos}
\end{figure*}

In this section we investigate the spectral characteristics of a quantum network in which
 each arm of the  Lieb-ladder geometry is `decorated' by diamond-shaped Aharonov-Bohm (AB) interferometer~\cite{aharony6}.
Each elementary interferometer is pierced by a
invariable magnetic perturbation applied perpendicular to the plane of the
interferometer, and traps a flux $\Phi$ (in unit of $\Phi_0 = hc/e$). 
This type of diamond based interferometers have been formerly studied as the minimal prototypes of bipartite structures having nodes with different coordination numbers, and representing 
a family of itinerant geometrically
frustrated electronic systems~\cite{kampf}-\cite{lopes2}.
We refer to Fig.~\ref{ifm-model}(a). 
A standard diamond-Lieb AB interferometer is shown
 pictorially there whereas Fig.~\ref{ifm-model}(b) demonstrates that each diamond loop can take a shape of a
 \textit{quantum ring} consisting of multiple lattice points.
Each arm of the diamond may be decorated by $N$ number of
atomic scatterers between the vertices, such that the total number of
single level quantum dots contained 
in a single interferometer is $4(N + 1)$. An uniform magnetic flux $\Phi$ may be allocated within
each loop, and the electron hopping is restricted to take the non-vanishing value for the nearest neighboring nodes only.

To study the systematic spectral analysis we take the help of RSRG approach. 
Each elementary loop of the interferometer is \textit{properly} renormalized to transform it into a simple
 diamond having just four sites. 
 Due to this decimation process we will get 
 three types of sites $A$, $B$ and $C$ (respectively 
 marked by black, red and blue colored atomic sites 
 in the Fig.~\ref{ifm-model}(a)) with corresponding parameters given by
\begin{eqnarray}
\tilde{\epsilon}_{A} &=& \epsilon + 6 t \frac{U_{N-1}(x)}{U_{N}(x)} \nonumber \\
\tilde{\epsilon}_{B} &=& \epsilon + 4 t \frac{U_{N-1}(x)}{U_{N}(x)} \nonumber \\
\tilde{\epsilon}_{C} &=& \epsilon + 2 t \frac{U_{N-1}(x)}{U_{N}(x)} \nonumber \\
t_{F(B)} &=& t e^{\pm i (N+1) \theta}/U_{N}(x)
\label{interfero-rg}
\end{eqnarray}
Here, $U_{N}(x)$ is the $N$-th order Chebyshev polynomial of second kind, and $x= (E - \epsilon)/2t$.
The `effective' diamond loops are then renormalized in a proper way ($C$ types of sites are being 
decimated out)
such that we will get back the Lieb ladder with renormalized on-site potential and overlap integral respectively given by
\begin{eqnarray}
\tilde{\epsilon_4} &=& \tilde{\epsilon_B}+ \frac{4 t_{F} t_{B}}{(E-\epsilon_C)} \nonumber\\
\tilde{\epsilon_6} &=& \tilde{\epsilon_A}+ \frac{6 t_{F} t_{B}}{(E-\epsilon_C)} \nonumber\\
\tilde{t} &=& \frac{2 t_{F} t_{B}}{(E-\epsilon_C)}
\end{eqnarray}
We will now exploit all the above equations to extract the detailed information
about the electronic spectrum and the nature of the eigenstates provided by such a model interferometer.

\subsection{Spectral landscape and inverse participation ratio}
To analyze we first put $N=0$ here so that the quantum ring of elementary interferometer takes the 
form of a diamond (Fig.~\ref{ifm-model}(a)).
The density of states with energy for different values of magnetic flux enclosed within each elementary interferometer is shown in the upper panel of the
 Fig.~\ref{ifm-dos}. From the plot, we see that in absence of magnetic flux the density of states reflects the periodic nature of the geometry. It consists of absolutely continuous zones populated by resonant eigenstates with sharp spikes at $E = 0$ and $\pm 2$. But here it is to be noted that the localized character of those modes cannot be distinctly revealed because of its position within the continuum of extended modes. But when we apply quarter flux quantum the central localized mode becomes isolated and prominent.
It is also seen from the plots that with the gradual increment of flux value the window of resonant modes in the DOS profile shrinks along the energy scale and ultimately leads to \textit{extreme localization} of eigenstates for half flux quantum. Actually, the effective overlap parameter between the two axial extremities of the interferometer vanishes for this special flux value and this makes the complete absence of resonant modes to be possible. This is the basic physical background of extreme localization of excitation. We should appreciate that this typical flux induced localization of wave train inside a \textit{charateristic trapping island} is 
a subset of the usual Aharonov-Bohm caging~\cite{vidal}

For the sake of completeness of the discussion related to the spectral property of such quantum interferometer model, we have also calculated the inverse participation ratio (IPR) to certify the above density of states plots. To formulate the localization of a normalized eigenstate the inverse participation ratio is 
defined as
\begin{equation}
I = \sum_{i=1}^{L} |\psi_{i}|^4
\label{inverse}
\end{equation}
It is known that for an extended mode IPR goes as $1/L$, but it approaches to unity for a localized state.
The lower panel of Fig.~\ref{ifm-dos} describes the variation of IPR with the energy of the injected projectile for different flux values. It is evident from the plots that the IPR supports the spectral profile cited in the upper panel of Fig.~\ref{ifm-dos}. As we see that with nominal strength of perturbation the central gap opens up around $E=0$, clearly indicating the central localized mode. The shrinking of resonant window with the gradual increment of flux is also apparent from the IPR plots. It is also interesting to appreciate that for half flux quantum IPR plot (Fig.~\ref{ifm-dos}f) also demonstrates the AB-caging leading to the extreme localization of eigenstates.

\subsection{Flux dependent eigenspectrum}

\begin{figure}[ht]
\centering
\includegraphics[clip,width=6 cm,angle=0]{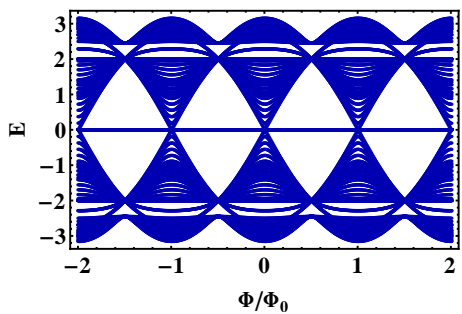}
\caption{(Color online) Flux dependent allowed eigenspectrum for the diamond-Lieb AB-interferometer
model. The pattern is flux periodic.}  
\label{ifm-spec}
\end{figure}
Fig.~\ref{ifm-spec} represents the essential graphical variation of allowed eigenspectrum for a diamond-Lieb AB interferometer with $N=0$ with respect to the external magnetic flux. With the increment of $N$, the number of scatterers in each elementary interferometer, the spectrum will be densely packed with several band crossings. The present variation is seen to be flux periodic of periodicity equal to one flux quantum. It is needless to say that the eigenspectrum is inevitably sensitive to the numerical values of the parameters of the Hamiltonian. However, the periodicity retains for such spectrum after every single flux quantum change of the external perturbation. 

It is observed that there is a tendency of clustering of the allowed eigenvalues towards
 the edges of the eigenspectrum as is clear from the above-mentioned diagram.
 A number of band crossings are noticed and the spectrum cites
 kind of a \textit{zero band gap semiconductor} like behavior, mimicking \textit{Dirac point} as observed in case of graphene, at $\Phi/\Phi_0 = \pm i$, $i$ being an integer including zero. As we increase the complexity in each interferometer by increasing $N$, the central gap gets
 consequently filled up by more eigenstates, and the $E-\Phi$ contours get
 more flattened up forming a quasi-continuous spectrum, exotic in nature. 
The central eigenstate corresponding to eigenvalue $E=0$ is a robust kind of mode irrespective of the application of perturbation, i.e., the existence of that state is insensitive to the value of the external flux. 
Moreover, when the magnetic flux is set as $\Phi = (i+1/2) \Phi_0$, we observe a spectral collapse. In that case one can easily identify the localization character of the central state.

 Most importantly, it is evident from the spectral landscape that the it consists of a set of discrete points (eigenvalues) for half flux quantum. This is the canonical case of \textit{extreme localization}. For such special flux value the vanishing overlap parameter makes the geometry equivalent to discrete set of lattice points with zero connectivity between them. This makes the excitation to be caged within the trapping island. Further it is to be noted that this AB-caging~\cite{vidal} may happen for any value of $N$, the number of eigenvalues in the discrete set depends on the choice of $N$.
 
\section{Closing remarks}
\label{close}
A methodical analysis of the flux induced tunable caging of excitation in a quasi-one dimensional Lieb network with long range connectivity is reported in this manuscript within the tight-binding framework. With the inclusion of second neighbor overlap integral in a decorated way, external source of perturbation can act as an important role for the selective caging of wave packet. Flux dependent band engineering and hence the comprehensive control over the group velocity of the wave train as well as the band curvature are studied in details.
Decoration of the next nearest neighbor hopping in certain quasiperiodic fashion or by some deterministic fractal object is also demonstrated analytically. Real space renormalization group approach provides us a suitable platform to obtain an exact prescription for the determination of self-localized modes induced by destructive quantum interference effect. As we have seen that in the quasiperiodic Fibonacci variation the distribution of eigenstates shows a standard three-subband pattern while in case of fractal entity countably infinite number of localized modes cite an interesting quasi-continuous distribution against flux.  We have also critically studied the spectral properties of a diamond Lieb interferometer. 
The energy spectrum shows an exotic feature comprising
extended, staggered and edge-localized eigenfunctions. The number of such states depend on the number of quantum dots present in each
arm of the elementary diamond interferometer, and can populate
the spectral landscape as densely as desired by the experimentalists.
A constant magnetic perturbation can be utilized to control the positions
of all such states. Moreover at special flux value the spectrum describes the \textit{Aharonov-Bohm caging} of eigenstates leading to an interesting spectral collapse. 


\begin{acknowledgments}
The author is thankful for the stimulating
discussions regarding the results with Dr. Amrita Mukherjee.
The author also gratefully acknowledges the fruitful discussion made with
Prof. A. Chakrabarti. 
\end{acknowledgments} 

\end{document}